\documentstyle[12pt]{article}

\newcommand{\be}{\begin{equation}}
\newcommand{\bea}{\begin{eqnarray}}
\newcommand{\ee}{\end{equation}}
\newcommand{\eea}{\end{eqnarray}}

\def\theequation{\arabic{section}.\arabic{equation}}
\begin{document}
\setcounter{equation}{0}
 \renewcommand{\theequation}{\arabic{equation}}
\topmargin -1cm \oddsidemargin=0.25cm\evensidemargin=0.25cm
\setcounter{page}1
\renewcommand{\thefootnote}{\fnsymbol{footnote}}
\begin{flushright}
LTH 882
\end{flushright}
\vskip .7in
\begin{center}
{\Large \bf Current Exchanges for Reducible Higher Spin Modes on AdS
\footnote{Talk given at the XIXth International Colloquium on Integrable Systems and Quantum Symmetries,
Prague, Czech Republic, June 17-19, 2010} }
\vskip .7in {\large Angelos  Fotopoulos$^a$}\footnote{e-mail:
{\tt foto@to.infn.it}}
and {\large Mirian Tsulaia$^b$}\footnote{e-mail: {{\tt tsulaia@liv.ac.uk}, Associate member 
of the Centre of Particle Physics and Cosmology, Ilia State University, 0162, Tbilisi Georgia }} \vskip .2in
{$^a$\it Dipartimento di Fisica Teorica dell'Universit\`a di Torino
and INFN\\Sezione di Torino,
via P. Giuria 1, I-10125 Torino, Italy} \\
\vskip .2in {$^b$ \it Department of Mathematical Sciences, University of Liverpool,
Liverpool, L69 7ZL, United Kingdom}\\

\end{center}
\vskip .3in
\begin{abstract}
We show how to decompose a Lagrangian for reducible massless bosonic Higher Spin modes into the 
ones describing irreducible (Fronsdal) Higher Spin modes on a ${\cal D}$ dimensional $AdS$ space. Using this decomposition
we construct a new Nonabelian cubic interaction vertex  
for reducible higher spin modes and two scalars on $AdS$ from the already known vertex which involves irreducible (Fronsdal) modes.

\end{abstract}


Higher Spin gauge theories   (see \cite{Vasiliev:2004qz}--\cite{Fotopoulos:2008ka} for recent reviews) are usually formulated
either in frame--like \cite{Fradkin:1986qy}--\cite{Bandos:1999qf}  or  metric -- like \cite{Fronsdal:1978rb}--\cite{Francia:2002pt}
 approaches.

Recently several interesting cubic vertexes have been constructed in the metric -- like approach \cite{Metsaev:1997nj}, \cite{Bekaert:2005jf}
\cite{Zinoviev:2007js}, \cite{Manvelyan:2010jr}.
Bearing in mind a possible application of the reducible Higher Spin modes (described by the so called ``triplet'' \cite{Francia:2002pt} )
 for String Theory \cite{Sagnotti:2010at}--\cite{Polyakov:2010qs} and for AdS/CFT correspondence \cite{Giombi:2009wh}
we consider the problem of cubic interaction of a triplet on  AdS space. In particular we study the cubic
interaction of a triplet 
with two scalar fields.

The main result of this paper is twofold. Firstly, we show that the procedure derived in \cite{Fotopoulos:2009iw} to decompose
the free Lagrangian for reducible massless bosonic Higher Spin modes in a flat space time works
for an arbitrary dimensional AdS space as well. The second and more important result is that after 
this decomposition one can use the cubic vertex\footnote{Let us point out that the method given in \cite{Buchbinder:2006eq} 
describes construction of {\it nonabelian} cubic interaction vertexes, see also \cite{Fotopoulos:2007yq} for some explicit {\it nonabelian} examples.
A particular example of an abelian vertex given in \cite{Fotopoulos:2007nm} is in some sense a ``degenerate'' solution of the
method, where however the abelian property is maintained in a nontrivial way, due to the structure of the  ghost terms. }
of \cite{Fotopoulos:2007yq}, which describes an interaction of  irreducible (Fronsdal) Higher Spin modes
with two scalars, to obtain an interaction vertex for reducible Higher Spin modes with two scalars.
Obviously this technique can be applied not only for a particular vertex given in \cite{Fotopoulos:2007yq},
but for construction of more complicated interaction vertexes in AdS following a method given in \cite{Buchbinder:2006eq}.
The advantage of this approach is that a construction of interaction vertexes for triplets
in AdS is often technically complicated due to repeated commutators between covariant derivatives 
and the double tracelessness condition for irreducible Higher Spin modes makes the problem at hand 
considerably simpler.

Let us start from a free Lagrangian describing the propagation of reducible massless Higher Spin modes
on a ${\cal D}$ dimensional AdS space.
It contains a field $\varphi_{\mu_1,...,\mu_s}(x)$ of the rank $s$, a field $C_{\mu_1,...,\mu_{s-1}}(x)$ of  rank $s-1$ and 
a field $D_{\mu_1,...,\mu_{s-2}}(x)$ of a rank $s-2$ and has the form \cite{Sagnotti:2003qa} (see also \cite{Fotopoulos:2008ka}
for the details of the construction),
\begin{eqnarray} \nonumber
{\cal L} &=& - \, \frac{1}{2}\ (\nabla_\mu \varphi)^2 \ + \ s\,
\nabla \cdot \varphi \, C \ + \ s(s-1)\, \nabla \cdot C \, D \
 + \ \frac{s(s-1)}{2} \, (\nabla_\mu D)^2 \ - \ \frac{s}{2} \,
C^2 \\ \nonumber &+& \ \frac{s(s-1)}{2L^2}\, {(\varphi^{'})}^2 \ - \
\frac{s(s-1)(s-2)(s-3)}{2L^2} \, {(D^{'})}^2
  \ - \ \frac{4s(s-1)}{L^2} \, D \, \varphi^{'} \\
&-& \ \frac{1}{2L^2} \, \left[ (s-2)({\cal D}+s-3) \, - \, s
\right] {\varphi}^2 \ + \ \frac{s(s-1)}{2L^2} \, \left[ s({\cal
D}+s-2)+6 \right]\,  D^2  , \label{LtripletBADS}
\end{eqnarray}
The
symbol $\nabla \cdot$ means divergence, while $\nabla$  is
symmetrized action of $\nabla_\mu$ on a tensor. The symbol
${}^\prime$ means that we take the trace of a field.
Multiplication of a tensor by the metric $g$ implies 
symmetrized multiplication, i.e., if $A$ is a vector $A_\mu$ we
have $g A = g_{( \mu \nu} A_{\rho)}= g_{\mu \nu} A_\rho +g_{\mu \rho} A_\nu+g_{\nu \rho}
A_\mu $.
This Lagrangian is invariant under the gauge transformations
with  parameter  $\Lambda_{\mu_1,...,\mu_{s-1}}(x)$
\begin{eqnarray}
&& \delta \varphi \ = \ \nabla \, \Lambda \ , \nonumber \\
&& \delta  C  \ = \ \Box \;
\Lambda + \frac{(s-1)(3-s-{\cal
D})}{L^2}\
  \Lambda
+ \frac{2}{L^2} \,  g \; \Lambda^\prime
 \nonumber \\
&& \delta D \ = \ \nabla \cdot
\Lambda \ . \label{adstripletgauge}
\end{eqnarray}
Let us note that the field $C(x)$ has no kinetic term and  can be eliminated via its own equations
of motion to obtain
\begin{eqnarray}\label{Lcomp2}
{\cal L} &=& - \, \frac{1}{2}\ (\nabla_\mu \varphi)^2 \ + \
\frac{s}{2} \,
(\nabla \cdot \varphi)^2 \ + \ s(s-1)\, \nabla \cdot \nabla \cdot \varphi \, D \ \nonumber \\
&+& \ s(s-1) \ (\nabla_\mu D)^2 \ + \ \frac{s(s-1)(s-2)}{2} \,
(\nabla \cdot D)^2  \\ \nonumber
&+& \frac{s(s-1)}{2L^2}\, {(\varphi^{'})}^2 \ - \
\frac{s(s-1)(s-2)(s-3)}{2L^2} \, {(D^{'})}^2
  \ - \ \frac{4s(s-1)}{L^2} \, D \, \varphi^{'} \\ \nonumber
&-& \ \frac{1}{2L^2} \, \left[ (s-2)({\cal D}+s-3) \, - \, s
\right] {\varphi}^2 \ + \ \frac{s(s-1)}{2L^2} \, \left[ s({\cal
D}+s-2)+6 \right]\,  D^2  .
\end{eqnarray}
Now we would like to decompose this Lagrangian in terms of irreducible (Fronsdal) \cite{Fronsdal:1978vb}
modes, following the procedure given in \cite{Fotopoulos:2009iw} for a  Minkowski space.

Let us start with the simplest example of a $s=2$ triplet which contains fields $\varphi_{\mu \nu}(x)$,
$C_\mu(x)$ and $D(x)$. Let us make the ansatz 
\begin{equation}
\varphi_{\mu \nu}= \Psi_{\mu \nu} + \frac{1}{{\cal D} -2} g_{\mu \nu} \Psi, \quad 
 \varphi^{'}-2D=\Psi.
\end{equation}
Inserting these expressions back to the Lagrangian (\ref{Lcomp2}) for $s=2$
one obtains
 \begin{eqnarray}\label{Ls2} \nonumber
 {\cal L}&=& -{1 \over 2} (\nabla_\mu \Psi_{\rho \sigma})^2 +
 (\nabla_\nu \Psi^\nu_\mu)^2 + \Psi^{'}
 \nabla_{\mu}\partial_{\nu} \Psi^{\mu \nu} + {1\over 2}
 (\nabla_\mu \Psi^{'})^2 - {1 \over 2({\cal D}-2)} (\nabla_\mu \Psi)^2 \\ 
&+& \frac{1}{L^2} (\Psi_{\mu \nu})^2 + \frac{{\cal D}-3}{L^2 ({\cal D}-2)} (\Psi)^2 + \frac{{\cal D}-3}{2L^2 } {(\Psi^{'})}^2
 \end{eqnarray}
Therefore, the initial Lagrangian (\ref{Lcomp2}) has been decomposed into a sum of two Fronsdal Lagrangians
for $s=2$ field $\Psi_{\mu \nu}$ with the gauge transformation law $\delta \Psi_{\mu \nu} = \nabla_\mu \Lambda_\nu +\nabla_\mu \Lambda_\nu $
 and a gauge invariant scalar $\Psi$.

Let us describe this procedure for the spin $4$ triplet, since in this case both a constraint on the parameter of gauge transformations
and an off--shell constraint on the gauge field arise. Let us use the substitution \cite{Fotopoulos:2009iw}
\begin{eqnarray}\label{phiDs4}
\varphi^{(4)} &=& \Psi^{(4)} + {1\over {\cal D}+2} g \Psi^{(2)} + {1\over {\cal D}({\cal D}-2)} {(g)}^2 \Psi^{(0)} \nonumber \\
D&=& {1\over 2} [ {\Psi'}^{(4)} +{2\over {\cal D}+2} \Psi^{(2)} + {1\over {\cal D}+2} g
{\Psi'}^{(2)} +{2\over {\cal D}({\cal D}-2)} g \Psi^{(0)}].
\end{eqnarray}
The field $\Psi^{(4)}$ is doubly traceless and transforms under the gauge transformations as
\begin{equation}
\delta \Psi^{(4)}= \nabla \tilde \Lambda,
 \quad \tilde \Lambda= \Lambda - {1\over {\cal D}+2}
 \eta \Lambda' 
\end{equation}
Inserting these expressions into the Lagrangian  (\ref{Lcomp2}) one can see again that it decomposes into the sum 
of Fronsdal modes with spins $4,2$ and $0$, described by the fields $ \Psi^{(4)},  \Psi^{(2)}$ and  $\Psi^{(0)}$.
 
One can further generalize this procedure for an arbitrary spin. In particular take
\begin{eqnarray}\label{fD}
\varphi &=&  \sum_{k=0}^{[{s \over 2}]}
\tilde{\rho}_k({\cal D},s) {(g)}^k \Psi^{(s-2k)} \nonumber \\
D &=& {1\over
2}\sum_{k=0}^{[{s \over 2}]-1} \tilde{\rho}_k({\cal D},s) {(g)}^k
\Psi^{'({s-2k})} + \sum_{k=1}^{[{s \over 2}]} \tilde{\rho}_k({\cal D},s)
{(g)}^{k-1} \Psi^{(s-2k)}.
\end{eqnarray}
and 
\begin{equation}\label{tLambda}
{\tilde \Lambda}_{s-1-2k}= \sum_{q=0}^{[{s\over 2}]}
\rho_q({\cal D},s-2k-1) {(g)}^q \Lambda^{[q+k]({s-1})}
\end{equation}
with
\begin{equation}\label{rho}
\rho_q({\cal D}-2,s)=  {(-1)^q
({\cal D}+2(s-q-3))!! \over ({\cal D}+2(s-3))!!}, \quad \tilde{\rho}_k({\cal D},s)= {({\cal D}+2(s-2k-2))!! \over ({\cal D}+2(s-k-2))!!}
\end{equation}
and $[q+k]$ denotes a number of traces.
Finally, one can show that the normalization factor
for propagators for each of individual Fronsdal mode
i.e., the
inverse of the prefactor of  $(\nabla_\mu \Psi^{(s-2k)})^2$ terms
multiplied by 2 is
\begin{equation}\label{Q}
Q(s,k,{\cal D})= { 2^k k! (s-2k)!\over s!{\tilde \rho}_k({\cal D},s)}.
\end{equation}
Now let us build a cubic interaction vertex of a Higher Spin triplet with two scalars on AdS. To this end let us use
the corresponding vertex for an  individual Fronsdal mode \cite{Fotopoulos:2007yq}
\begin{equation} \label{INV}
{\cal L}_{int}^{00s}= \Psi^{(s)} \cdot J_s + [\frac{s-1}{6L^2} [2s^2 + (3 {\cal D}-4)s -6] - \frac{s-2}{L^2}]\Psi^{' (s)} \cdot J_{s-2} 
\end{equation}
where
\begin{equation}
J^{1;2}_{s-2q}=\sum_{r=0}^{s-2q}  C_{s-2q}^{r}  (-1)^r 
(\nabla^{\mu_1} \dots
\nabla^{\mu_r}
\phi_1)\ (\nabla^{\mu_{r+1}} \dots
\nabla^{\mu_{s-2q}}\phi_2)\
\end{equation}
Therefore  multiplying    interacting vertexes (\ref{INV}) with the appropriate factor
(\ref{Q}) and adding them to the free Lagrangian (\ref{Lcomp2}), one finds the expression for a  cubic Lagrangian
describing the interaction of reducible Higher Spin modes with two scalars on AdS.
After that one can perform a current- current exchange procedure following the lines of
\cite{Francia:2007qt}, \cite{Bekaert:2009ud}, \cite{Fotopoulos:2009iw}.

\noindent {\bf Acknowledgements.}
It is a pleasure to thank A.P. Isaev, S.O. Krivonos and A.O. Sutulin
for valuable discussions.
The work of A. F. was supported by
an INFN postdoctoral fellowship and partly supported by the Italian
MIUR-PRIN
contract 20075ATT78.
The work of M.T. has been supported by a STFC rolling grant ST/G00062X/1.

\end{document}